\RequirePackage{fix-cm}
\documentclass[final]{svjour3}
\smartqed  
\usepackage{graphicx}
\usepackage{rotating}
\usepackage{amssymb}
\usepackage{amsmath}
\usepackage{mathptmx}
\usepackage[numbers]{natbib}
\usepackage{bm}
\journalname{Journal of Low Temperature Physics}
\begin{document}
\title {The optical conductivity for a spin-Peierls ground state of (TMTTF)$_{2}$PF$_{6}$ with tetramer formation }
\author{ T. Yamaguchi  \and K. Iwano }
\institute{Institute of Materials Structure Science, High Energy Accelerator Research Organization (KEK), 1-1 Oho, 
Tsukuba 305-0801, Japan \\ Tel.: +81-29-864-1171 \\ \email{tymgc@post.kek.jp} }
\date{\today / Accepted: date}
\maketitle
\begin{abstract}
We theoretically investigate the optical conductivity of (TMTTF)$_{2}$PF$_{6}$ in the spin-Peierls ground state 
within the framework of the exact diagonalization method at absolute zero temperature ($T=0$). As an effective model, 
a $1/4$-filled 1D (one-dimensional) extended Hubbard model with tetramerization is employed. Using appropriate parameters 
of the model which have already been reported, we clarify the electronic photoexcitation energies from the spin-Peierls ground state. 
Since some experiments indicate the formation of a tetramer in the spin-Peierls ground state of (TMTTF)$_{2}$PF$_{6}$, 
our results are useful to understand the effects of tetramerization on the optical properties of (TMTTF)$_{2}$PF$_{6}$. 
\end{abstract}
%
%
%
%
%
\section{Introduction}
\label{sec1}
\par
A quasi-1D organic conductor (TMTTF)$_{2}$X (TMTTF = tetramethyltetrathiafulvalene, X = anion) 
which is one of the Fabre charge-transfer salts possesses various physical phases and has been actively studied 
\cite{G1,G2,G3,G4,G5,G6,G7,Ogata,G8}. 
The minimal model of such materials has been treated as a 1/4-filled hole or a 3/4-filled electron system of a 1D extended 
Hubbard model with dimerization for many years because of the fact that nearest two TMTTF molecules constitute a dimer. 
However, observations for intramolecular vibrations of TMTTF molecules by means of the Raman spectroscopy 
have suggested that the nearest two dimers may also form a tetramer in low-temperature phases of (TMTTF)$_{2}$X type 
compounds \cite{Tetr}. 
In particular, a tetramer formation at 7 K has recently been reported by the detailed X-ray structural analysis 
of (TMTTF)$_{2}$PF$_{6}$ \cite{Sawa}. Here, according to the Ref. \cite{G2}, (TMTTF)$_{2}$PF$_{6}$ exhibit a charge-ordered phase 
below 67 K and a spin-Peierls phase below 19 K. Besides, the charge-ordered state is maintained even in 
that spin-Peierls phase \cite{Tetr,Sawa}. 
\par
To investigate optical properties of various physical phases in (TMTTF)$_{2}$PF$_{6}$, while optical conductivities have been 
measured \cite{ODexp,ODexp2}, they are poor temperature dependences. In addition, a plasmalike reflectivity edge 
peculiar to a metal state has been reported recently in the charge-ordered insulator phase of (TMTTF)$_{2}$AsF$_{6}$ \cite{Iwai} 
which is one of the similar substances of (TMTTF)$_{2}$PF$_{6}$. 
Due to above situations, it is extremely difficult to extract the information of pure electronic excitations from 
observed optical conductivities. 
\par
In this article, we theoretically calculate the optical conductivity for the spin-Peierls ground state of 
(TMTTF)$_{2}$PF$_{6}$ with tetramer formation by using the exact diagonalization method at $T=0$ and reveal characteristics 
of electronic excitation energies. Throughout this article, we take $\hbar=e=1$ and the lattice constant equals unity for simplicity. 
\section{Formulation}
\par
\begin{figure*}[t]
  \begin{center}
    \includegraphics[width=5cm,keepaspectratio]{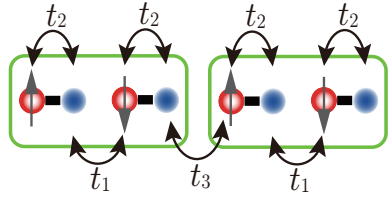}
    \caption{Schematic picture for the spin-Peierls ground state of (TMTTF)$_{2}$PF$_{6}$ in a 1D chain. 4 dimers, 2 tetramers, 
                and all related transfer integrals are illustrated. Circles and up (down) arrows on them are molecular orbitals of TMTTF and 
                up (down) spins, respectively. A dimer is represented as two circles combined with a flat bar. Neighboring two dimers 
                configure a tetramer which is displayed in the region surrounded by a box. Due to exhibiting charge-orders, 
                a charge rich site and a charge poor site alternate. }
    \label{fig1}
  \end{center}
\end{figure*}
\par
\par
We consider a 1D chain of $N_{s}$ sites based on 
a 1/4-filled hole system with an equal population of spins ($N_{\uparrow}=N_{\downarrow}=N_{s}/4$) at $T=0$. 
Our Hamiltonian with the PBC (periodic boundary condition) is described as 
\begin{equation}
H = -\sum_{j=1}^{N_{s}}\sum_{\sigma}t(j)\left[
c_{j+1,\sigma}^{\dagger}c_{j,\sigma} + c_{j,\sigma}^{\dagger}c_{j+1,\sigma}
\right] + U\sum_{j=1}^{N_{s}}n_{j,\uparrow}n_{j,\downarrow} + V\sum_{j=1}^{N_{s}}n_{j+1}n_{j}, 
\label{eq1}
\end{equation} 
where $c_{j,\sigma}^{(\dagger)}$ denotes an annihilation (creation) operator of a hole with spin $\sigma=\uparrow, \downarrow$ 
at the $j$-th site and $n_{j}\equiv n_{j,\uparrow} + n_{j,\downarrow}$ ($n_{j,\sigma}\equiv c_{j,\sigma}^{\dagger}c_{j,\sigma}$). 
A tetramer formation of (TMTTF)$_{2}$PF$_{6}$ in the spin-Peierls ground state is classified by utilizing 
different transfer integrals defined as 
\begin{equation}
t(j) = \begin{cases}
t_{1} & {\rm for} \; j=4l-2 \; \textrm{(inter-dimer), } \\
t_{2} & {\rm for} \; j=4l-1, 4l-3 \; \textrm{(intra-dimer), } \\
t_{3} & {\rm for} \; j=4l \; \textrm{(inter-tetramer), }
\end{cases}
\label{eq2}
\end{equation}
for $1\leq l\leq N_{s}/4$, respectively. The relationship between these transfer integrals and the ground state 
are schematically illustrated in 
Fig. \ref{fig1}. 
According to the Ref. \cite{Sawa}, $t_{1}/t_{2}=0.862$ and $t_{3}/t_{2}=0.833$ are calculated within the framework of the 
extended H\"{u}ckel method \cite{Huckel} with structural parameters of (TMTTF)$_{2}$PF$_{6}$ observed by X-ray diffraction 
experiments at 7 K. In contrast to transfer integrals, to determine Coulomb repulsive interaction strengths $U$ and $V$ 
is much difficult in general. However, we employ $0.2\leq V/U\leq 0.6$ for $U/t_{2}=5, 10$ as typical values of (TMTTF)$_{2}$PF$_{6}$ 
\cite{Ogata,Para1,Para2,Para3,Para4,Para5,Para6} in this article. 
\par
Considering a weak photoexcitation where the linear response theory is legitimated, an optical conductivity of given photon energy 
$\omega>0$ is written as 
\begin{equation}
\sigma(\omega) = -\frac{1}{N_{s}\omega}{\rm Im}\left[
\langle\psi_{0}|J\frac{1}{\omega+i\eta+E_{0}-H}J|\psi_{0}\rangle
\right] \; (\eta\rightarrow 0+),  
\label{eq3}
\end{equation}
where $J  = i\sum_{j=1}^{N_{s}}\sum_{\sigma}t(j)[
c_{j+1,\sigma}^{\dagger}c_{j,\sigma} - c_{j,\sigma}^{\dagger}c_{j+1,\sigma}
]$ represents the electrical current operator. Here, $|\psi_{0}\rangle$ is the ground state wavefunction of $H$ in Eq. (\ref{eq1}) and 
$|\psi_{0}\rangle$ is calculated by means of the exact diagonalization method with its energy $E_{0}$. 
\par
For the following discussions, we derive free dispersions of $H$ in Eq. (\ref{eq1}) ($U=V=0$) for the thermodynamic limit 
($N_{s}\rightarrow+\infty$). Using $\alpha, \alpha^{\prime}=\pm$, they have the forms, 
\begin{equation}
E_{\alpha,\alpha^{\prime}}(k) = \alpha\sqrt{
\frac{t_{1}^{2}+2t_{2}^{2}+t_{3}^{2}}{2} + \alpha^{\prime}\sqrt{
\left[\frac{(t_{1}+t_{3})^{2}}{4} + t_{2}^{2}
\right](t_{1}-t_{3})^{2} +4t_{1}t_{2}^{2}t_{3}\cos^{2}(2k)
}
}. 
\label{eq4}
\end{equation} 
The first Brillouin zone of these dispersions is $-k_{\rm F}\leq k<k_{\rm F}$, where $k_{\rm F}=\pi/4$ denotes a Fermi wave number 
corresponding to a 1/4-filling. 
\section{Optical conductivities and electronic excitation energies}
\par
\begin{figure*}[t]
  \begin{center}
    \includegraphics[width=12.2cm,keepaspectratio]{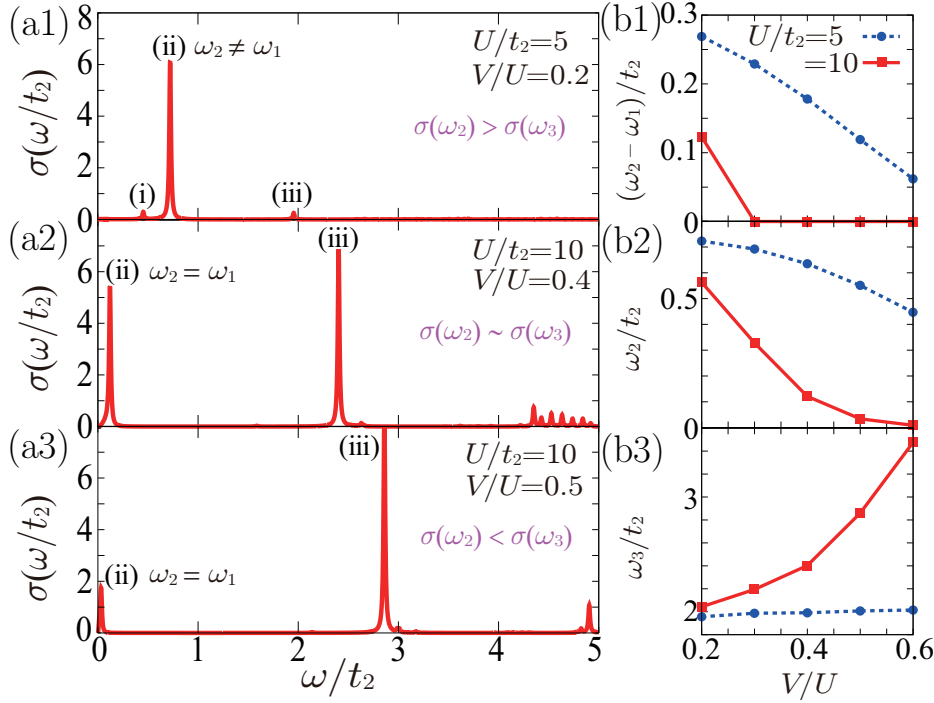}
    \caption{(a1)-(a3) Calculated $\sigma(\omega)$ for $N_{s}=20, \eta/t_{2}=0.01$ under the PBC. Using the well-known order of 
    $t_{2}\sim 0.2$ eV \cite{Ogata,Para4}, our displayed $\omega$ region is $0< \omega \lesssim$ 1 eV which corresponds to a 
    typical $\omega$ range of observations. Three characteristic excitation energies of electrons in the $\omega$ region 
    are illustrated as (\romannumeral1) $\omega_{1}$, (\romannumeral2) $\omega_{2}$, and (\romannumeral3) $\omega_{3}$ in the 
    figures. 
    $U, V$ dependences of $\omega_{2}-\omega_{1}$, $\omega_{2}$, and $\omega_{3}$ are shown in (b1), (b2), and (b3), respectively. 
    }
    \label{fig2}
  \end{center}
\end{figure*}
\par
%
%
%
%
\par
\begin{figure*}[t]
  \begin{center}
    \includegraphics[width=10.7cm,keepaspectratio]{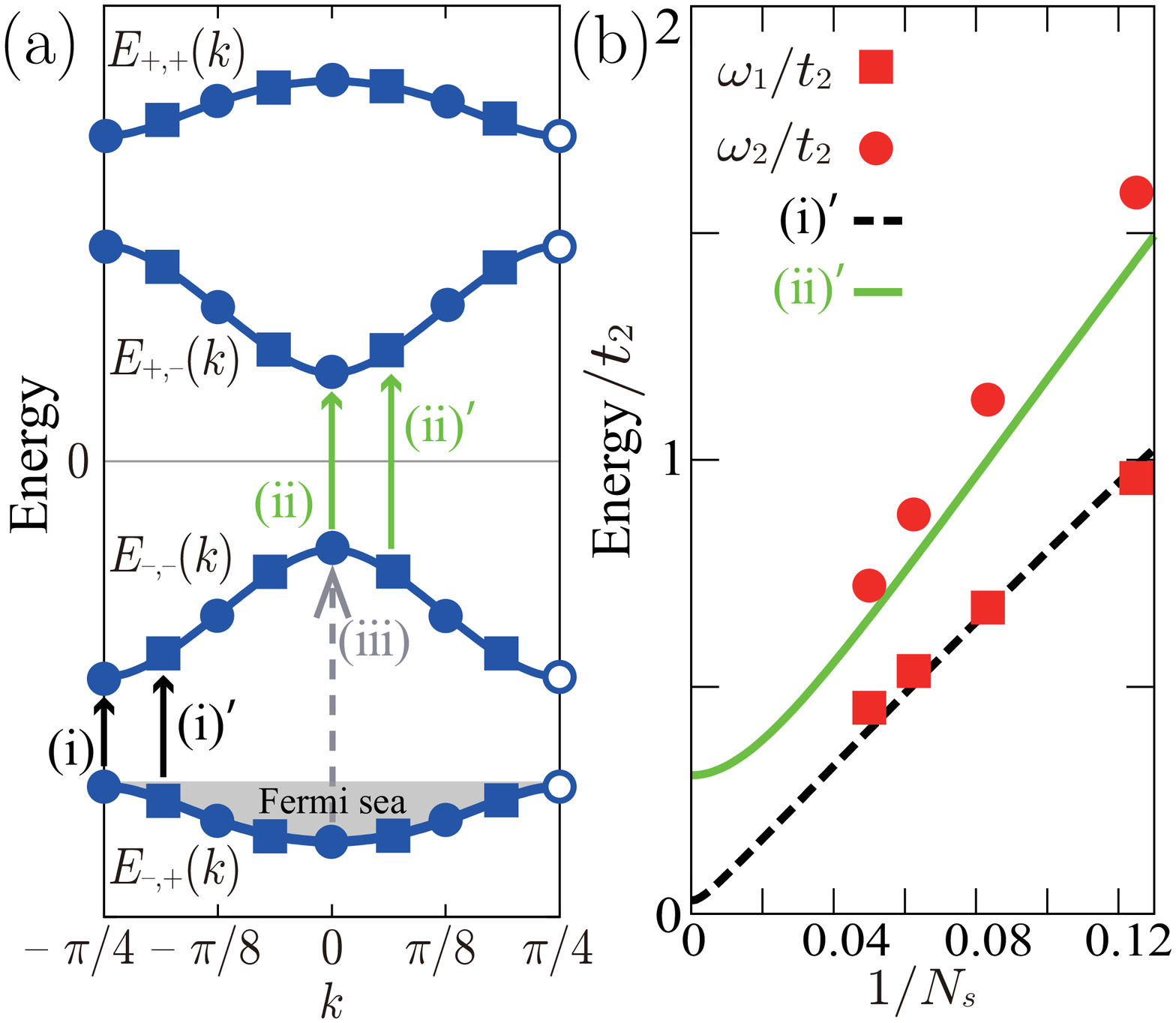}
    \caption{(a) Schematic picture of free dispersions for $N_{s}=16$ on the first Brillouin zone. Solid lines are 
    $E_{\alpha,\alpha^{\prime}}(k)$ in Eq. (\ref{eq4}). Filled circles and squares represent allowed discrete wave numbers of 
    the PBC and those of the APBC, respectively, for $N_{s}=16$. Hollowed circles are unavailable discrete wave numbers. 
    (\romannumeral1) is a vertical transition at $k=\pm\pi/4=\pm k_{\rm F}$ and (\romannumeral1)$^{\prime}$ is its nearest transition. 
    All corresponding discrete wave numbers of the transition (\romannumeral1)$^{\prime}$ are 
    $k_{n}^{\rm (\romannumeral1)^{\prime}}=\pm(\pi/4-\pi/N_{s})$ both for $N_{s}=8, 16$ under the APBC and 
    $N_{s}=12, 20$ under the PBC. (\romannumeral2) and (\romannumeral3) are transitions at $k=0$. 
    (\romannumeral2) also corresponds to the minimum energy gap in the spinless fermion picture. To evaluate the finite size effects of 
    (\romannumeral2), we choose (\romannumeral2)$^{\prime}$ as its nearest transition. All corresponding discrete 
    wave numbers of the transition (\romannumeral2)$^{\prime}$ are $k_{n}^{\rm (\romannumeral2)^{\prime}}=\pm\pi/N_{s}$ and 
    $k_{n}^{\rm (\romannumeral2)^{\prime}}$ are, however, only valid for the APBC. 
    (b) Finite size scalings of $\omega_{1}, \omega_{2}$ for $N_{s}=8, 16$ under the APBC, $N_{s}=12, 20$ under the PBC, and 
    $U/t_{2}=5, V/U=0.2$. (\romannumeral1)$^{\prime}$ and (\romannumeral2)$^{\prime}$ are the same as in (a). The dashed line and 
    the solid line express $\Delta E_{\rm (\romannumeral1)^{\prime}}(N_{s}) = E_{-,-}(k_{n}^{\rm (\romannumeral1)^{\prime}}) - 
    E_{-,+}(k_{n}^{\rm (\romannumeral1)^{\prime}})$ and $\Delta E_{\rm (\romannumeral2)^{\prime}}(N_{s}) = 
    E_{+,-}(k_{n}^{\rm (\romannumeral2)^{\prime}}) - E_{-,-}(k_{n}^{\rm (\romannumeral2)^{\prime}})$, respectively, where 
    $E_{\alpha,\alpha^{\prime}}(k)$ are in Eq. (\ref{eq4}). For the conventional dimerized model (in Eq. (\ref{eq1}) with $t_{3}=t_{1}$), 
    electronic excitation energies of the transitions (\romannumeral1) and (\romannumeral2) are corresponding to 0 (gapless) and 
    $\Delta_{\rm L}^{\rm di}=2|t_{1}-t_{2}|$, respectively. 
    }
    \label{fig3}
  \end{center}
\end{figure*}
\par
\par
Typical results of optical conductivities $\sigma(\omega)$ with $\eta/t_{2}=0.01$ are shown in 
Figs. \ref{fig2} (a1)-(a3) 
and we find that three significant peaks in the low-energy region represented as (\romannumeral1), (\romannumeral2), 
and (\romannumeral3) in the figures characterize $\sigma(\omega)$. Here, we note that our calculations are performed 
with $N_{s}=20$ for the computational problem although 
finite size effects remain quintessentially in the order of $1/N_{s}$. 
Now, we introduce corresponding electronic excitation energies $\omega_{1}$, $\omega_{2}$, and $\omega_{3}$ of the peaks 
(\romannumeral1), (\romannumeral2), and (\romannumeral3), respectively ($\omega_{1}\leq \omega_{2}<\omega_{3}$). 
Using this, we first investigate $U, V$ dependences of $\sigma(\omega)$ as shown in 
Figs. \ref{fig2} (b1)-(b3). 
As a result, we can classify the structures of $\sigma(\omega)$ into two types. 
One type is the case of $\omega_{1}\neq\omega_{2}$ and $\sigma(\omega_{2})>\sigma(\omega_{3})$ 
seen in $U/t_{2}=5$ with $0.2\leq V/U\leq 0.6$ and in $U/t_{2}=10$ with $0.2\leq V/U\lesssim 0.3$. 
A distinctive $\sigma(\omega)$ of this case is shown in 
Fig. \ref{fig1} (a1). 
Another type is the case of $\omega_{1}=\omega_{2}$ for $U/t_{2}=10$ with $0.3\lesssim V/U\leq 0.6$ and 
typical results of $\sigma(\omega)$ are shown in 
Figs. \ref{fig2} (a2) and (a3). 
In this case, $\sigma(\omega_{2})> \sigma(\omega_{3})$ for $0.3\lesssim V/U\lesssim 0.4$, 
$\sigma(\omega_{2})\sim \sigma(\omega_{3})$ for $V/U\sim 0.4$, 
and otherwise $\sigma(\omega_{2})< \sigma(\omega_{3})$ are satisfied. 
\par
From 
Fig. \ref{fig2} (b3), 
$\omega_{3}\rightarrow 2t_{2}$ might be fulfilled for $V\rightarrow 0$ with fixed $U$ or for $U\rightarrow +\infty$ with fixed $V$. 
Furthermore, $\omega_{3}$ enlarges with increase in $V$. This leads us to judge $\omega_{3}$ as an electronic excitation energy of a COI 
(charge-ordered insulator) state originates from $U,V$. According to the phase diagram of the conventional dimerized model 
(in Eq. (\ref{eq1}) with $t_{3}=t_{1}$) at $T=0$, the ground state can be divided into a dimer-Mott insulator phase 
for small $U,V$ and a COI phase for large $U,V$ \cite{Ogata,Para3}. As mentioned in Sect. \ref{sec1}, nature of the spin-Peierls phase of 
(TMTTF)$_{2}$PF$_{6}$ partially contains that of the COI phase. In addition to this, the critical point of 
the metal-COI phase transition is $(U, V)=(+\infty,2t)$ for a 1/4-filled extended Hubbard model (in Eq. (\ref{eq1}) with 
$t_{1}=t_{2}=t_{3}\equiv t$) \cite{14exHubP}. Then, the growth of $\omega_{3}$ with respect to finite 
$V$ for $U\rightarrow +\infty$ can roughly be estimated by $V-2t_{2}$ or, namely, $\omega_{3}\propto V$ and that origin might be 
related to the COI phase. This feature certainly appears in 
Fig. \ref{fig2} (b3), 
especially, for $U/t_{2}=10$ (in the COI phase). 
\par
On the other hand, as seen in 
Figs. \ref{fig2} (b1) and (b2), 
we cannot apply above discussions of $\omega_{3}$ to $U, V$ dependences of $\omega_{1}$ and $\omega_{2}$. However, for 
the conventional dimerized model (in Eq. (\ref{eq1}) with $t_{3}=t_{1}$), peak structures of $\sigma(\omega)$ in the low-energy region 
have already been manifested within the framework of the exact diagonalization method \cite{Mila}. According to 
the Ref. \cite{Mila}, there are two specific excitation energies of electrons 
$\omega_{1}^{\rm di}\equiv\Delta_{\rm L}^{\rm di}=2|t_{1}-t_{2}|$ and 
$\omega_{2}^{\rm di}\equiv\Delta_{\rm F}^{\rm di}=2\sqrt{t_{1}^{2}+t_{2}^{2}}$ ($\omega_{1}^{\rm di}<\omega_{2}^{\rm di}$) which are 
almost independent in small $U,V$ (not in the COI phase). 
$\Delta_{\rm L}^{\rm di}$ and $\Delta_{\rm F}^{\rm di}$ are corresponding to vertical transitions of 
free dispersions at the zone-boundary of the first Brillouin zone and at the Fermi surface in that model, respectively. 
We note that the transition of $\Delta_{\rm L}^{\rm di}$ is permitted for the spinless fermion picture which is, for instance, 
valid for $U\rightarrow+\infty$ and $V=0$ \cite{OgataShiba}. 
Using this as a reference, we inquire into finite size scalings with $U/t_{2}=5, V/U=0.2$ (not in the COI phase or, in other word, 
in the regime of weak interactions) which is the minimum parameter set in our calculations and try to grasp the connection between 
electronic excitation energies ($\omega_{1}$, $\omega_{2}$, $\omega_{3}$) and free dispersions in Eq. (\ref{eq4}). 
For this purpose, all calculations of the finite size scalings are done with the APBC 
(the anti-periodic boundary condition) for $N_{s}=8$, 16 and the PBC for $N_{s}=12$, 20 due to avoiding forbidden 
electronic excitations at the zone-boundaries of dispersions in the first Brillouin zone \cite{Mila}. Here, under the APBC, 
the first term on the right side of Eq. (\ref{eq1}) is just treated as 
$-\sum_{j=1}^{N_{s}-1}\sum_{\sigma}t(j)[
c_{j+1,\sigma}^{\dagger}c_{j,\sigma} + c_{j,\sigma}^{\dagger}c_{j+1,\sigma}]+t(N_{s})[
c_{N_{s}+1,\sigma}^{\dagger}c_{N_{s},\sigma} + c_{N_{s},\sigma}^{\dagger}c_{N_{s}+1,\sigma}
]$. 
\par
Schematic pictures of electronic excitations associated with free dispersions in Eq. (\ref{eq4}) and the results of the finite size scalings 
are shown in Figs. \ref{fig3} (a) and (b), respectively. Vertical transitions represented as (\romannumeral1), (\romannumeral2) and 
(\romannumeral3) in 
Fig. \ref{fig3} (a) 
are the same as in Fig. \ref{fig2}. 
Deducing from 
Fig. \ref{fig3} (b) 
and explanations in the caption of 
Fig. \ref{fig3}, 
$\omega_{1}$ and $\omega_{2}$ are good agreement with 
$\Delta E_{\rm (\romannumeral1)^{\prime}}(N_{s})/t_{2}$ and $\Delta E_{\rm (\romannumeral2)^{\prime}}(N_{s})/t_{2}$, respectively. 
Then, $\omega_{1}$ in the thermodynamic limit seems to converge on 
$\Delta_{\rm F}^{\rm tetra}\equiv\Delta E_{\rm (\romannumeral1)^{\prime}}
(N_{s}\rightarrow+\infty)=E_{-,-}(\pm k_{\rm F}) - E_{-,+}(\pm k_{\rm F})=0.029t_{2}\ll \omega_{1}^{\rm di}=0.276t_{2}$. Here, 
$\Delta_{\rm F}^{\rm tetra}$ denotes the minimum band gap represented as (\romannumeral1) in 
Fig. \ref{fig3} (a) 
and corresponds to the inter-band transition at the Fermi surface for $N_{s}\rightarrow +\infty$. In a similar fashion, 
$\omega_{2}$ in the thermodynamic limit seems to converge on 
$\Delta_{\rm L}^{\rm tetra}\equiv\Delta E_{\rm (\romannumeral2)^{\prime}}
(N_{s}\rightarrow+\infty)=E_{+,-}(0) - E_{-,-}(0)=0.305t_{2}$ expressed as (\romannumeral2) in 
Fig. \ref{fig3} (a) 
and $\Delta_{\rm L}^{\rm tetra}\sim\omega_{1}^{\rm di}$. This means that, due to $t_{1}\sim t_{3}$, the minimum inter-band gap energy 
in the spinless fermion picture of our tetrameric model is close to that of the conventional dimerized model 
(in Eq. (\ref{eq1}) with $t_{3}=t_{1}$). 
Contrary to the above-discussed case of large $U, V$ (strong interactions), $\omega_{3}$ hardly depends on $V$ for $U/t_{2}=5$ 
as shown in 
Fig. \ref{fig2} (b3) 
and the value $\omega_{3}=1.956t_{2}$ at $U/t_{2}=5$ and $V/U=0.2$ is comparable to 
$\Delta_{\rm U}^{\rm tetra}\equiv E_{-,-}(0) - E_{-,+}(0)=1.695t_{2}\sim \omega_{2}^{\rm di}=1.320t_{2}$. 
Here, $\Delta_{\rm U}^{\rm tetra}$ corresponds to the inter-band transition (\romannumeral3) illustrated in 
Fig. \ref{fig3} (a). 
However, we note that this transition does not physically correspond to the transition of $\omega_{2}^{\rm di}$ for 
the conventional dimerized model (in Eq. (\ref{eq1}) with $t_{3}=t_{1}$). 
\par
Consequently, in the thermodynamic limit, our results indicate that optical conductivities with tetramer formation are characterized 
by three excitation energies of electrons $\omega_{1}\sim\Delta_{\rm F}^{\rm tetra}$, $\omega_{2}\sim\Delta_{\rm L}^{\rm tetra}$, and 
$\omega_{3}\sim\Delta_{\rm U}^{\rm tetra}$ ($\Delta_{\rm F}^{\rm tetra}< \Delta_{\rm L}^{\rm tetra}< \Delta_{\rm U}^{\rm tetra}$) 
for not in the COI phase (or weak Coulomb interactions) such as the case of $U/t_{2}=5$ and $V/U\sim 0.2$. 
On the other hand, for strong Coulomb interactions like 
$U/t_{2}=10$, $V/U\sim 0.6$ (in the COI phase), $\omega_{1}=\omega_{2}\ll \Delta_{\rm L}^{\rm tetra}$ and 
$\Delta_{\rm U}^{\rm tetra}\ll \omega_{3}\propto V$ are satisfied. The feature of $\omega_{3}\propto V$ can be regarded as 
the similar case of the COI phase with the well-known conventional model (in Eq. (\ref{eq1}) with $t_{1}=t_{2}=t_{3}\equiv t$). 
However, the detailed evaluation of $\omega_{1}=\omega_{2}$ ($\ll \Delta_{\rm L}^{\rm tetra}$) in the thermodynamic limit 
is far difficult due to the strong correlations caused by large $U, V$. 
\section{Conclusion}
\par
In summary, we theoretically calculate the optical conductivity of (TMTTF)$_{2}$PF$_{6}$ in the spin-Peierls ground state 
within the framework of the exact diagonalization method at $T=0$. For computations, we treat a $1/4$-filled 1D extended 
Hubbard model with tetramerization and appropriate parameters which have already been reported. As a result, we clarified that 
the electronic excitation energies from that spin-Peierls ground state are characterized by $\Delta_{\rm F}^{\rm tetra}$, 
$\Delta_{\rm L}^{\rm tetra}$, and $\Delta_{\rm U}^{\rm tetra}$ 
($\Delta_{\rm F}^{\rm tetra}< \Delta_{\rm L}^{\rm tetra}< \Delta_{\rm U}^{\rm tetra}$) for weak Coulomb interaction strengths 
(not in the COI phase). From comparison with the results of the conventional dimerized model, the tetramerization newly 
produces the electronic excitation energy $\Delta_{\rm F}^{\rm tetra}$ which is the lowest gap energy at the Fermi surface on the 
free dispersions. 
This can be an instrumental feature which is presented in the optical conductivity to distinguish electronic excitations of dimers from 
those of tetramers in the low-energy region. However, for strong Coulomb interactions (in the COI phase), apart from 
the excitation energy which is roughly proportional to $V$, strong correlations drastically affect electronic excitation energies 
even in the low-energy region and it is hard to evaluate them. 
Although calculations in this article contain finite size effects to some extent, our results are still useful to understand the 
effects of tetramerization on the optical properties of (TMTTF)$_{2}$PF$_{6}$. 
\begin{acknowledgements}
This study was supported by the Grant-in-Aid for Scientific Research from JSPS (Grant No. 17K05509) and 
JST CREST Grant No. JPMJCR1661. 
\end{acknowledgements}

\end{document}